\documentclass[english,aps,fleqn,pra,twocolumn,showpacs]{revtex4-1}
\usepackage[T1]{fontenc}
\usepackage[latin9]{inputenc}
\usepackage{color}
\usepackage{babel}
\usepackage{amsmath}
\usepackage[unicode=true,pdfusetitle,
 bookmarks=true,bookmarksnumbered=false,bookmarksopen=false,
 breaklinks=false,pdfborder={0 0 1},backref=false,colorlinks=false]
 {hyperref}
\usepackage{breakurl}

\makeatletter

\@ifundefined{textcolor}{}
{%
 \definecolor{BLACK}{gray}{0}
 \definecolor{WHITE}{gray}{1}
 \definecolor{RED}{rgb}{1,0,0}
 \definecolor{GREEN}{rgb}{0,1,0}
 \definecolor{BLUE}{rgb}{0,0,1}
 \definecolor{CYAN}{cmyk}{1,0,0,0}
 \definecolor{MAGENTA}{cmyk}{0,1,0,0}
 \definecolor{YELLOW}{cmyk}{0,0,1,0}
}

\makeatother

\begin{document}

\title{Effective Subspace for Pure State Transfer Induced by Measurements}

\author{Yaoxiong Wang$^{1,2}$, Fang Gao$^{1}$, Xubing Tang$^{3}$, Feng
Shuang$^{1,2}$}

\affiliation{$^{1}$Institute of Intelligent Machines, Chinese Academy of Sciences,
Hefei, China 230031\\
$^{2}$Department of Automation, University of Science and Technology
of China, Hefei, 230026, China\\
$^{3}$Department of Mathematics \& Physics, Anhui University of Technology,
Ma'anshan 243032, China}
\begin{abstract}
{\normalsize{}One main goal of quantum control is to steer a quantum
system toward an expected state or dynamics. For measurement-induced
quantum control, measurements serve as the only control, which is
like the cases in quantum Zeno and anti-zeno effects. In this paper,
this scenario will be investigated in a general $N$-level quantum
system. It is proved that, when the intial and expected states are
both pure, the control space could be reduced to an effective subspace
spanned by these two states only. This result will greatly simplify
the measurement-induced control strategy of an $N$-level quantum
system.}{\normalsize \par}
\end{abstract}


\maketitle

\section{introduction}

Quantum control has drawn much attention both theoretically and experimentally\cite{Rabitz2000,shapiro2003principles,d2007introduction,Petersen2010,Pechen2010}
due to its potential applications. The goal of quautum control is
usually to steer a quantum system into an expected quantum state via
external fields\cite{Huang1983}. However, the control means are not
limited to introducing control fields. With rapid development of the
research field, various approaches were proposed to play the role
of control\cite{Pechen2010}. The quantum measurement, as one of the
necessary criteria for a quantum computer, does not only serve as
the read-out but also could be employed as the control\cite{wiseman2009quantum}. 

The idea of utilizing measurements to drive quantum systems origins
from the famous quantum Zeno effect which claims that repeated frequent
measurement of an unstable state will keep it unchanged\cite{Misra1977,Itano1990}.
The counterpart of Zeno effect, namely anti-Zeno effect , was also
discussed\cite{Balachandran2000,Lewenstein2000}, and both of them
were observed in an unstable system\cite{Fischer2001}. In contrast
to control field schemes, measurement-based quantum control will introduce
irreversible decoherence inevitably. By combining with coherent control,
it can be applied to closed-loop molecular states control\cite{Judson1992},
quantum dynamics engineering\cite{Lloyd2001}, dephasing decoherence
control\cite{Zhang2007} and remote state preparation\cite{Mandilara2005}.
Controllability analysis of quantum systems shows that measurements
could transform a non complelely controllable system into a controllable
one while unitary evolution only could not achieve this task\cite{VilelaMendes2003,Sugny2008}.
System controllability under Kraus-map, which includes non-projective
measurements, was also studied\cite{Wu2007,Pechen2008}. More extreme
cases with measurement-only control schemes have been studied. \textcolor{black}{In
the absence of active coherent controls, projective measurements were
introduced to map an unknown mixed state onto a known expected pure
state while only two noncommuting observables are available\cite{Roa2006}.
Optimized measurement sequence was obtained for two-level quantum
systems\cite{Pechen2006,Shuang2008}. }The continuous measurement
extension was investigated\textcolor{black}{{} and it shows that quantum
anti-Zeno effect could be recovered in the limit of continuous cases}\cite{Shuang2007,Shuang2008,Tarn2009,Jacobs2010,Petersen2010}.
Schemes for state engineering using measurements and fixed dynamics
were also peoposed\cite{Pedersen2014}. Recently, a novel quantum
state preparing protocol was presented in which only a restricted
set of measurements is required\cite{Ashhab2010}. In this protocol,
previous measurement outcome is feedback as the control, and it shows
that arbitrary state could indeed be prepared. This feedback idea
was also introduced to state manipulation\cite{Fu2014}.

\textcolor{black}{These above works investigated both two-level}\cite{Pechen2006,Zhang2007,Shuang2008,Pechen2008,Ashhab2010}\textcolor{black}{{}
and }$N$\textcolor{black}{-level systems\cite{Lloyd2001,VilelaMendes2003,Roa2006,Wu2007,Pechen2010,Wang2010,Pedersen2014}.
In this paper, we will investigate how to control }$N$\textcolor{black}{-level
systems with arbitrary projective measurements only, which is a direct
extension of our previous works\cite{Wang2010}. However, powerful
tools in two-level context, i.e. Pauli matrics and Stokes vector,
are no longer valid in $N$-level systems with $N>2$. In Ref\cite{Wang2010},
a variational method over unitary group was introduced to obtain the
necessary condition for optimal measurement sequence. This condition
was carefully analyzed for pure initial and expected states with a
single measurement. This will be extended to arbitrary finite times
of projective measurements for pure state-to-state control. We find
that the conrol space will be reduced to an }effective\textcolor{black}{{}
subspace spanned by the initial and expected states only. The transition
cases through intermedia levels are excluded and the two-level model\cite{Pechen2006,Shuang2008}
will be recovered.}

\textcolor{black}{The rest of the paper is organized as follows. In
Sec. II, we introduce some basic notations, the variational method
over unitary group and the necessary condition for an optimal measurement
sequence under non-selective assumption. In Sec. III, we investigate
an orthogonal state-to-state control case, and derive the }effective
subspace\textcolor{black}{{} for optimal control. These cases will be
generalized to ones of non-orthognonal pure states in Sec. IV. }Discussion
and conclusion\textcolor{black}{{} are presented in Sec.V.}

\section{\textcolor{black}{necessary} condition for optimal state transfer}

The measurement-based optimal control case for a two-level quantum
system was solved by using Pauli matrices, and the optimal projectors
were also obtained\cite{Pechen2006}. Since this approach seems invalid
for general $N$\textcolor{black}{-level} systems, we introduce a
virational method and derive a necessary condition for the optimal
\textcolor{black}{measurement sequence}\cite{Wang2010}\textcolor{black}{.
In this section, some basic concepts and notations are firstly be
introduced. Then, the virational method and the necessary condition
for the optimal measurement sequence, which appears as a chain of
equalities, are presented . These tools will be utilized }to study$N$\textcolor{black}{-level
cases in the next section.}

\textcolor{black}{Consideing an }$N$\textcolor{black}{-level quantum
system whose states is characterized by a density matrix }$\rho$\textcolor{black}{,
the state after a non-selective measurement }$\mathcal{M}$\textcolor{black}{{}
becomes }$\mathcal{M}\left(\rho\right)=\sum_{i=1}^{N}P_{i}\rho P_{i}$\textcolor{black}{{}
where }$\{P_{i}\}$ \textcolor{black}{is a set of rank-1 projectors
satisfying the equations }$P_{i}P_{j}=\delta_{ij}P_{i}$\textcolor{black}{{}
and }$\sum_{1}^{N}P_{i}=I$\textcolor{black}{. Because of these properties,
we could always find a suitable unitary matrix }$U$\textcolor{black}{{}
to diagonalize these projectors, i.e. }$P_{i}=U\left|i\right\rangle \left\langle i\right|U^{\dagger}$
for $i=1,\cdots,N$, where $\left|i\right\rangle $ indicates a specific
level of the system. With these notations, the \textcolor{black}{non-selective}
measurement process could be recast as
\begin{equation}
\mathcal{M}\left(U\right)\rho=\sum_{i=1}^{N}U\left|i\right\rangle \left\langle i\right|U^{\dagger}\rho U\left|i\right\rangle \left\langle i\right|U^{\dagger}\label{eq:1}
\end{equation}
which contains three successive operations of $\rho$: (1) rotate
the basis (a unitrary transfermation) (2) set non-diagonal elements
to be zero (3) rotate the basis inversely.

Limited by the measurement techniques nowadays, it is difficult to
performe too many projective measurements within finite time interval.
Assuming that at most $m$-times measurement could be employed in
this interval,\textcolor{black}{{} the final state }$\rho^{m}$\textcolor{black}{{}
after these measurements reads }$\rho^{m}=\mathcal{M}(U_{m})\cdots\mathcal{M}(U_{1})\rho$.
We neglect free evolution of the system because this could be included
via a picture transformation\cite{Pechen2006,Shuang2008}. The control
objective $J$, which quantifies this control effect, is defined as
\textcolor{black}{the overlap between }$\rho^{m}$ \textcolor{black}{and
some expected final state }$\theta$:
\begin{equation}
J=Tr\left(\rho^{m}\theta\right)=Tr\bigl((\mathcal{M}(U_{m})\cdots\mathcal{M}(U_{1})\rho)\theta\bigr).\label{eq:2}
\end{equation}

The necessary condition for optimal control sequence is obtained by
considering the first derivative of $J$ with respect to variables
$\{U_{i}\}$. A variational analysis method is introduced to \textcolor{black}{parametrize}
the \textcolor{black}{neighborhood} of a unitary transformation and
the necessary condition\textcolor{black}{{} presents }as \textcolor{black}{a
chain of equalities:}

\begin{align}
 & [\rho,\mathcal{M}(U_{1})...\mathcal{M}(U_{m})\theta]\nonumber \\
= & \cdots\nonumber \\
= & [\mathcal{M}(U_{k})...\mathcal{M}(U_{1})\rho,\mathcal{M}(U_{k+1})...\mathcal{M}(U_{m})\theta]\nonumber \\
= & \cdots\nonumber \\
= & [\mathcal{M}(U_{m})...\mathcal{M}(U_{1})\rho,\theta].\label{eq:3}
\end{align}
Detailed mathematical derivation of the above equalities and some
special cases for single measurement was shown in \cite{Wang2010},
but general cases with $m$-times measurements are still intractable.
The rest of this paper will focus on this general scenario.

\section{effective subspace for optimal orthogonal pure states transfer}

In this section, the effective subspace for orthogonal pure state-to-state
optimal measurement control will be investigated. Assuming that the
initial and expected states are $\rho=\left|1\right\rangle \left\langle 1\right|$
and $\theta=\left|2\right\rangle \left\langle 2\right|$, respectively,
and all states are represented in the basis $\{\left|i\right\rangle ,i=1,2,....N\}$.
Due to the special form of $\rho$, $[\rho,\mathcal{M}(U_{1})...\mathcal{M}(U_{m})\theta]$
becomes a matrix with non-zero elements only in the first row and
first column . By the same reason, $[\mathcal{M}(U_{m})...\mathcal{M}(U_{1})\rho,\theta]$
becomes a matrix with non-zero elements only in the second row and
second column. Combining these two forms, we have

\begin{align}
 & [\rho,\mathcal{M}(U_{1})...\mathcal{M}(U_{m})\theta]\nonumber \\
= & \cdots\nonumber \\
= & [\mathcal{M}(U_{m})...\mathcal{M}(U_{1})\rho,\theta]\nonumber \\
= & \begin{bmatrix}\begin{array}{cc}
0 & c\\
-c^{*} & 0
\end{array} & O\\
O & O
\end{bmatrix}=C\label{eq:4}
\end{align}
where $c$ is a complex number and $O$ is a zero matrix of suitable
size. The case of $c=0$ is dropped henceforth, since it corresponds
to a trivial control which can not be the optimal one.

To clarify the discussion below, some notations are introduced here:
$I^{n}$ denotes $n\times n$ identity matrix; $U(n)$ denotes $n$-dimensional
unitary group and $U^{n}$ represents some elements in $U(n)$; $O^{n}$
is $n\times n$ zero matrix; $D(n)$ is the set of $n\times n$ density
matrices, i.e. positive semidefinite Hermitian matrices. With these
notations, we will first investigate the symmetry property of $\{U_{i}\}$
and propose lemma.1.

\textbf{Lemma.1 }if $\{U_{1},U_{2},\ldots,U_{m}\}$ is an optimal
control sequence, then $\{U_{s}U_{1},U_{s}U_{2},\ldots,U_{s}U_{m}\}$
could also achieve this optimum, where $U_{s}=e^{i\psi_{1}}\oplus e^{i\psi_{2}}\oplus U_{arb}$.
Here, $\psi_{1}$ , $\psi_{2}$ are arbitrary phases and $U_{arb}\in U(N-2)$.

The proof of lemma.1 is straightforward. Assuming that $\{U_{s}^{1}U_{1},U_{s}^{2}U_{2},\ldots,U_{s}^{m}U_{m}\}$
could also achieve this optimum, then we have sufficient conditions
for optimal control: (i)$U_{s}^{l\dagger}U_{s}^{l+1}=I$ for $l=1,\ldots,m-1$
and (ii) $\rho$ and $\theta$ are invariant under the action of $U_{s}^{1}$
and $U_{s}^{m}$, respectively. Consequently, we find that $U_{s}^{1}=\cdots=U_{s}^{m}$
and $U_{s}^{1}=U_{s}^{m}=e^{i\psi_{1}}\oplus e^{i\psi_{2}}\oplus U_{arb}$. 

According to lemma.1, we could always find proper $\psi_{1}$ or $\psi_{2}$
to eliminate the phase in $c$ and make it real. Thus, $c$ will be
treated as a real number hereafter. 

Now, we investigate the first two parts of Eq.\eqref{eq:4} and get
\begin{eqnarray}
[\rho,\mathcal{M}(U_{1})\tau] & = & C\label{eq:5}\\{}
[\mathcal{M}(U_{1})\rho,\tau] & = & C\label{eq:6}
\end{eqnarray}
in which $\tau$ is defined as $\mathcal{M}(U_{2})...\mathcal{M}(U_{m})\theta$.
Multipling $U_{1}^{\dagger}$ and $U_{1}$ from left and right in
Eqs.\eqref{eq:5} and \eqref{eq:6}, and taking the $(p,q)$ matrix
element of both sides , we have
\begin{eqnarray}
 &  & (U_{1})_{1p}^{*}(U_{1})_{1q}\left((U_{1}^{\dagger}\tau U_{1})_{qq}-(U_{1}^{\dagger}\tau U_{1})_{pp}\right)\nonumber \\
 & = & \left((U_{1})_{1p}^{*}(U_{1})_{2q}-(U_{1})_{2p}^{*}(U_{1})_{1q}\right)c\label{eq:7}\\
 &  & (U_{1}^{\dagger}\tau U_{1})_{pq}\left((U_{1})_{1p}^{*}(U_{1})_{1p}-(U_{1})_{1q}^{*}(U_{1})_{1q}\right)\nonumber \\
 & = & \left((U_{1})_{1p}^{*}(U_{1})_{2q}-(U_{1})_{2p}^{*}(U_{1})_{1q}\right)c\label{eq:8}
\end{eqnarray}
Obviously, Eq.\eqref{eq:7} implies that if $(U_{1})_{1p}\neq0$ and
$(U_{1})_{1q}\neq0$ then $(U_{1}^{\dagger}\tau U_{1})_{qq}-(U_{1}^{\dagger}\tau U_{1})_{pp}$
could be expressed by $\{(U_{1})_{1i}\}$ and $\{(U_{1})_{2i}\}$
explicitly. Further, we can write the objective $J$ as $Tr(\mathcal{M}(U_{1})\rho\tau)=\sum_{i}(U_{1})_{1i}(U_{1})_{1i}^{*}(U_{1}^{\dagger}\tau U_{1})_{ii}$.
Note that $\{(U_{1}^{\dagger}\tau U_{1})_{ii}\}$ could be reconstructed
by $\{(U_{1})_{1i}\}$ and $\{(U_{1})_{2i}\}$ up to a constant, and
if we fix this constant, which corresponds to solving Eq.\eqref{eq:7}
and Eq.\eqref{eq:8} in a special case, then $(U_{1}^{\dagger}\tau U_{1})_{ii}$
will only depend on $\{(U_{1})_{1i}\}$ and $\{(U_{1})_{2i}\}$, and
so does $J$. This observation will give us the explicit form of $\tau$. 

Using lemma.1, we could diagonize the submatrix spanned by the last
$N-2$ rows and columns in $\tau$, and then investigate $(U_{1}^{\dagger}\tau U_{1})_{11}$.
Since $(U_{1}^{\dagger}\tau U_{1})_{11}$ only depends on $\{(U_{1})_{1i}\}$
and $\{(U_{1})_{2i}\}$, we have $\frac{\partial(U_{1}^{\dagger}\tau U_{1})_{11}}{\partial(U_{1})_{31}}=\frac{\partial\sum_{ij}(U_{1})_{i1}^{*}(U_{1})_{j1}\tau_{ij}}{\partial(U_{1})_{31}}=0$.
Again, by lemma.1, we can assign some $U_{arb}$ to make $\{(U_{1})_{i1}\}$
real for $i=3,\ldots,N$. This will simplify our calculation below.
Remember that there is a normalization constraint, i.e. $\sum_{i}(U_{1})_{i1}(U_{1})_{i1}^{*}=1$,
which implies that $\{(U_{1})_{i1}\}$ are not independent in $(U_{1}^{\dagger}\tau U_{1})_{11}$.
Due this constraint, we treat $(U_{1})_{N1}$ as a non-independent
variable and have
\begin{align}
 & \frac{\partial\sum_{ij}(U_{1})_{i1}^{*}(U_{1})_{j1}\tau_{ij}}{\partial(U_{1})_{31}}\nonumber \\
= & \frac{\partial\left((U_{1})_{11}\tau_{31}+(U_{1})_{21}\tau_{32}+c.c.\right)(U_{1})_{31}}{\partial(U_{1})_{31}}\nonumber \\
+ & \frac{\partial\left((U_{1})_{11}\tau_{N1}+(U_{1})_{21}\tau_{N2}+c.c.\right)(U_{1})_{N1}}{\partial(U_{1})_{31}}\nonumber \\
+ & \frac{\partial\left((U_{1})_{31}^{*}(U_{1})_{31}\tau_{33}+(U_{1})_{N1}^{*}(U_{1})_{N1}\tau_{NN}\right)}{\partial(U_{1})_{31}}\nonumber \\
= & 0\label{eq:9} 
\end{align}
Since $(U_{1})_{11},(U_{1})_{11}^{*},(U_{1})_{21},(U_{1})_{21}^{*}$
are indepentdent, three parts in Eq.\eqref{eq:9} gives $\tau_{13}=\tau_{31}=\tau_{23}=\tau_{32}=0$,
$\tau_{N1}=\tau_{N2}=\tau_{1N}=\tau_{2N}=0$ and $\tau_{33}=\tau_{NN}$.
By the same reason, we also have $\tau_{1i}=\tau_{2i}=0$ and $\tau_{ii}=\tau_{NN}$
if we investigate $\frac{\partial(U_{1}^{\dagger}\tau U_{1})_{11}}{\partial(U_{1})_{i1}}$
for $i=3,\ldots,N$. This implies $\tau\in D(2)\oplus dI^{N-2}$ where
$d$ is a normalized factor .

Next, we start from this special form of $\tau$ to explore what kind
of $\{U_{i}\}$ could achieve the optimum. We define $\tau=\tau_{s}\oplus dI^{N-2}$
where $\tau_{s}\in D(2)$ and write $\mathcal{M}(U_{1})\rho$ in a
block form, i.e. $\mathcal{M}(U_{1})\rho=\left[\begin{array}{cc}
A_{11} & A_{12}\\
A_{21} & A_{22}
\end{array}\right]$, in which $A_{11}$ and $A_{22}$ are $2\times2$ and $(N-2)\times(N-2)$
matrices respectively. According to Eq.\eqref{eq:6}, we get $(dI^{2}-\tau_{s})A_{12}=O$
whose solutions have two different branches: (a) $\tau_{s}$ has no
eigenvalue of $d$ or (b) $\tau_{s}$ has an eigenvalue of $d$, and
these two branches indicate $A_{12}$ is zero or not, respectively. 

For case (a), we have $A_{12}=O$ and $\mathcal{M}(U_{1})\rho=\left[\begin{array}{cc}
A_{11} & O\\
O & A_{22}
\end{array}\right]$. To acquire this form of $\mathcal{M}(U_{1})\rho$, $U_{1}$ must
be (a.1) $U_{1}\in\left(U(2)\oplus U(N-2)\right)\times\left(1\oplus U(N-1)\right)$
or (a.2) $U_{1}\in U(2)\oplus U(N-2)$. Note that these two different
cases origin from whether $A_{11}$ and $A_{22}$ have degenerated
eigenvalue. 

For (a.1), we define $U_{1}=V_{1}W_{1}$ in which $V_{1}\in U(2)\oplus U(N-2)$
and $W_{1}\in1\oplus U(N-1)$. Since the rotation $V_{1}$ makes $\rho$
and $\tau$ belong to $D(2)\oplus O^{N-2}$ and $D(2)\oplus dI^{N-2}$
respectively, a further rotation $W_{1}$ makes some diagonal elements,
except the first one, in $V_{1}^{\dagger}\rho V_{1}$ equal, which
are the degenerated eigenvalue in $A_{11}$ and $A_{22}$. $W_{1}$
will also make the diagonal elements in the correaspnding subspace
of $V_{1}^{\dagger}\tau V_{1}$ equal. This is due to the fact that
this two subspaces, i.e. $diag(a,0,\ldots,0)$ and $diag(b,d,\ldots,d)$
are essentially of the same form of $\alpha diag(a,0,\ldots,0)+\beta diag(1,1,\ldots,1)$
which will have same diagonal elements under the act of $W_{1}$.
This property also implies that $W_{1}$ must act on the whole $(N-1)$-dimensional
subspace to achieve the maximum. Define the rotation with this property
by $U_{e}(N)$, we get $U_{1}\in\left(U(2)\oplus U(N-2)\right)\times\left(1\oplus U_{e}(N-1)\right)$.
Now, we have $U_{2}\in U(2)\oplus U(N-2)$ from (a) and $U_{1}\in\left(U(2)\oplus U(N-2)\right)\times\left(1\oplus U_{e}(N-1)\right)$.
This implies $\mathcal{M}(U_{2})M(U_{1})\rho\in D(2)\oplus d'I^{N-2}$
where no degenerated eigenvalue exists between these two subspaces.
According to Eq. \eqref{eq:4}, it indicates that $\mathcal{M}(U_{3})...\mathcal{M}(U_{m})\theta\in D(2)\oplus dI^{N-2}$.
Again, we have $U_{3}$ belongs to $U(2)\oplus U(N-2)$ or $\left(U(2)\oplus U(N-2)\right)\times\left(1\oplus U_{e}(N-1)\right)$.
The latter one, i.e. $U_{3}\in\left(U(2)\oplus U(N-2)\right)\times\left(1\oplus U_{e}(N-1)\right)$,
implies that two eigenvalues in $\mathcal{M}(U_{2})...\mathcal{M}(U_{m})\theta$,
which are not $d$, are both larger or smaller than $d$. If they
are both larger than $d$, then $U_{1}$ will belong to $U(2)\oplus U(N-2)$
but not $\left(U(2)\oplus U(N-2)\right)\times\left(1\oplus U_{e}(N-1)\right)$.
If they are both smaller than $d$, then the objective could not exceed
$d$ which is less than $1/2$, and this is contrary to the known
optimum\textcolor{black}{\cite{Pechen2006,Shuang2008}}. Thus, we
have $U_{3}\in U(2)\oplus U(N-2)$. This procedure could be operated
in an iterative fashion, and we finally have $U_{i}\in U(2)\oplus U(N-2)$
for $i=2,\dots,N$. Note that if we choose $U_{1}=V_{1}$ and discard
$W_{1}$, then $J$ will be larger than the one when we have $U_{1}=V_{1}\times W_{1}$.
Thus we have $U_{i}\in U(2)\oplus U(N-2)$ for $i=1,\dots,N$.

If we have (a.2), then both $U_{1}$ and $U_{2}$ belong to $U(2)\oplus U(N-2)$,
and Eq.\eqref{eq:4} implies that $\mathcal{M}(U_{3})...\mathcal{M}(U_{m})\theta\in D(2)\oplus D(N-2)$.
By considering the special form of $\tau$, we also have $\mathcal{M}(U_{3})...\mathcal{M}(U_{m})\theta\in D(2)\oplus dI^{N-2}$
which indicates that $U_{3}\in U(2)\oplus U(N-2)$ or $U_{3}\in\left(U(2)\oplus U(N-2)\right)\times\left(1\oplus U(N-1)\right)$.
For the former one, we could consequently investigate $U_{4}$ which
revives the discussion of (a.2). And the later one is just case(b)
when we treat $U_{3}$ in (a.2) as $U_{2}$ in case (b). We will also
show that $U_{1}$ in case(b) also belongs to $U(2)\oplus U(N-2)$
as $U_{2}$ in (a.2). So, this situasion, i.e. (a.2), is equivalent
to case(b).

For case (b), the optimal control between $\tau$ and $\rho$ is equivalent
to the optimal control between two non-orthogonal pure states. This
could be seen when we diagonalize $\tau$ and get $diag[a,d,\ldots,d]$.
The solution of this case is obtained\textcolor{black}{\cite{Wang2010}}
and we have $U_{1}\in U(2)\oplus U(N-2)$. Define $U_{2}=V_{2}W_{2}$
where $V_{2}\in U(2)\oplus U(N-2)$, $W_{2}\in1\oplus U_{e}(N-1)$,
and
\begin{eqnarray*}
\chi & = & V_{2}^{\dagger}\mathcal{M}(U_{3})...\mathcal{M}(U_{m})\theta V_{2}\\
 & = & \left[\begin{array}{cc}
B_{11} & \begin{array}{c}
B_{21,up}\\
B_{21,down}
\end{array}\\
\begin{array}{cc}
B_{21,up}^{\dagger} & B_{21,down}^{\dagger}\end{array} & B_{22}
\end{array}\right]
\end{eqnarray*}
where $B_{21,up}$ and $B_{21,down}$ are $(N-2)$ dimensional row
vectors. Since arbitrary $U\in I^{2}\oplus U(N-2)$ will not change
the diagonal elements in $V_{2}^{\dagger}\mathcal{M}(U_{1})\rho V_{2}$
and a successive $W_{2}$ makes diagonal elements, except the first
one, in $V_{2}^{\dagger}\mathcal{M}(U_{1})\rho V_{2}$ equal, we find
that $B_{22}=dI^{N-2}$ and $B_{21,down}=0$. Define $\kappa=\sum_{i=1}^{N}W_{2}\left|i\right\rangle \left\langle i\right|W_{2}^{\dagger}V_{2}^{\dagger}\mathcal{M}(U_{1})\rho V_{2}W_{2}\left|i\right\rangle \left\langle i\right|W_{2}^{\dagger}$,
and Eq.\eqref{eq:4} gives $\kappa\chi-\chi\kappa=V_{2}^{\dagger}CV_{2}$
which implies $B_{21,up}=0$. Thus we have $\mathcal{M}(U_{3})...\mathcal{M}(U_{m})\theta\in D(2)\oplus d''I^{N-2}$
and again, $U_{3}\in U(2)\oplus U(N-2)$ or $U_{3}\in\left(U(2)\oplus U(N-2)\right)\times\left(1\oplus U_{e}(N-1)\right)$.
The latter one should be discarded since it will not achieve the maximum
when $U_{2}\in\left(U(2)\oplus U(N-2)\right)\times\left(1\oplus U(N-1)\right)$,
and finally we have $U_{3}\in U(2)\oplus U(N-2)$. Following the analysis
of (a.1) above, we also have $U_{i}\in U(2)\oplus U(N-2)$ for $i=1,\dots,N$.

To conclude, for orthogonal pure states control induced by projective
measurements only, the control space will be reduced to a two-dimensional
effective subspace which spanned by the initial and expected states
only.

\section{generalization to non-orthogonal pure states }

The analysis in Sec.III could be generalized to cases with non-orthogonal
pure states, and Eq.\eqref{eq:4} is still valid. Define these two
pure states by $\rho=\left|1\right\rangle \left\langle 1\right|$
and $\theta=\left(\alpha\left|1\right\rangle +\beta\left|2\right\rangle \right)\left(\alpha^{*}\left\langle 1\right|+\beta^{*}\left\langle 2\right|\right)$.
Since $\rho=\left|1\right\rangle \left\langle 1\right|$, we have
$[\rho,\mathcal{M}(U_{1})...\mathcal{M}(U_{m})\theta]=\begin{bmatrix}0 & c & s\\
c^{*} & 0 & O\\
s^{\dagger} & O & O
\end{bmatrix}$ where $c$ is a complex number and $s$ is a $(N-2)$-dimensional
row vector. On the other hand, if we asuume that $\theta$ could be
diagonalized by $u$, i.e. $u\theta u^{\dagger}=\left|1\right\rangle \left\langle 1\right|$,
then
\begin{eqnarray}
 &  & (u\oplus I^{N-2})\begin{bmatrix}0 & c & s\\
c^{*} & 0 & O\\
s^{\dagger} & O & O
\end{bmatrix}(u^{\dagger}\oplus I^{N-2})\nonumber \\
 & = & (u\oplus I^{N-2})[\mathcal{M}(U_{m})...\mathcal{M}(U_{1})\rho,\theta](u^{\dagger}\oplus I^{N-2})\nonumber \\
 & = & [\mathcal{M}((u\oplus I^{N-2})U_{m})...\mathcal{M}((u\oplus I^{N-2})U_{1})\nonumber \\
 &  & \bigl((u\oplus I^{N-2})\rho(u^{\dagger}\oplus I^{N-2})\bigr),\left|1\right\rangle \left\langle 1\right|]\label{eq:10}
\end{eqnarray}
The right hand of Eq.\eqref{eq:10} indicates that it will be a matrix
with non-zero elements in the first row and column only, thus $(u\oplus I^{N-2})\begin{bmatrix}0 & c & s\\
c^{*} & 0 & O\\
s^{\dagger} & O & O
\end{bmatrix}(u^{\dagger}\oplus I^{N-2})=\begin{bmatrix}0 & c' & s'\\
c'^{*} & 0 & O\\
s'^{\dagger} & O & O
\end{bmatrix}$. Since $u$ is nontrivial, we must have $s=O$ and Eq.\eqref{eq:4}
is recovered. The rest of the proof for non-orthogonal pure states
is the same as that in Sec.III. We could also conclude that the effective
subspace will be reduced to the one spanned by $\left|1\right\rangle $
and $\left|2\right\rangle $.

\section{discussion and conclusion}

In summary, we discuss the use of projective measurements to control
an $N$-level quantum system. Previous works of two-level scenario
are generalized assuming that the measurement time is limited. It
is proved that for pure initial and expected states, the control space
will be reduced to an effective subspace spanned by these two states.
The intuition for this result is straightforward, no extra level is
needed for optimal control. The situation for mixed states is much
complicated. Numerical calculations show that this conclusion still
works if ``pure states'' are replaced by ``mixed states'', and
its rigorous proof will be further investigated.
\begin{acknowledgments}
This work is supported by the National Natural Science Foundation
of China under Grant Nos. 61403362, 61473199, 61203061, 61374091 and 11574295.
\end{acknowledgments}

\bibliographystyle{apsrev}

\end{document}